\documentclass[pra,twocolumn,showpacs,preprintnumbers,amsmath,amssymb,superscriptaddress]{revtex4}

\usepackage{graphicx}
\usepackage{times}
\usepackage{graphicx}
\usepackage{dcolumn}
\usepackage{bm}
\newcommand{\be}{\begin{equation}}
\newcommand{\ee}{\end{equation}}
\newcommand{\ba}{\begin{array}}
\newcommand{\ea}{\end{array}}
\newcommand{\bea}{\begin{eqnarray}}
\newcommand{\eea}{\end{eqnarray}}

\newcommand{\ket}{\rangle}
\newcommand{\mbf}{\mathbf}

\begin{document}

\title{Quantum computation with doped silicon cavities}

\author{M. Abanto}\altaffiliation[Present address: ] {Centro de Ci\^encias Biol\'ogicas e da Natureza, Universidade Federal do Acre --
Caixa Postal 500, Rio Branco, AC 69915-900, Brazil}
\author{L. Davidovich}\author{ Belita Koiller}\author{ R. L. de Matos Filho}\affiliation{Instituto de F\'{\i}sica,
Universidade Federal do Rio de Janeiro, \\Caixa Postal 68528, Rio de
Janeiro, RJ 21941-972, Brazil}

\date{\today}






\begin{abstract}
 We propose a quantum computer architecture involving  subs\-titutional donors in photonic-crystal silicon cavities and the optical initialization, manipulation, and detection processes already demonstrated in ion traps and other atomic systems. Our scheme considerably simplifies the implementation of the building blocks for the successful operation of silicon-based solid-state quantum computers, including positioning of the donors, realization of one- and two-qubit gates, initialization and readout of the qubits.  Detailed consideration of the processes involved, using state-of-the-art values for the relevant parameters, indicates that this architecture might lead to errors per gate compatible with scalable quantum computation.
 \end{abstract}

\maketitle

The search for a working quantum computer has comprised areas
ranging from optics to atomic and condensed-matter
physics~\cite{nielsen}. Finding physical systems that allow for
accurate operations has been a formidable
challenge, yet to be met. Indeed, the viability of quantum computers depends on finding physical
systems that allow scalable fault-tolerant computation, which means
that the errors remain bounded when the number of qubits increases. In
order to have scalable quantum computation, the error per gate (EPG)
should be smaller than a certain threshold, which depends on the specific error-correction scheme. For independent and identically distributed errors, the best lower bound so far is $1.9\times 10^{-4}$ \cite{aliferis}. For other architectures, which require however a large resource overhead \cite{knill}, this threshold is bounded below by $1.04\times10^{-3}$ \cite{preskill}.

 Semiconductor
devices~\cite{Loss,Imamoglu}, and most particularly those based on
silicon~\cite{Kane}, have attracted considerable attention,
but actual realization is hindered by difficulties
concerning scalability, detection and fabrication~\cite{Kane05}.
Most candidates for a semiconductor-based quantum computer rely on
spin-1/2 fermion qubits~\cite{Petta}, which for Si may be associated to
the long-lived electron and nuclear spins of shallow substitutional
donors~\cite{Kane}. In fact, electronic spin decoherence times have been shown to be larger than 60 ms   in $^{28}$Si (isotopically
purified) at temperatures of 7 K~\cite{tspin}. Implementation of quantum computation with these systems  is hindered by several problems.
The most obvious~\cite{Kane05} is the difficulty in the manipulation
and measurement of single-spin states. Two-qubit operations
relying on exchange gates~\cite{Loss}, restricted to
nearest-neighbor interactions, have limited
scalability potential~\cite{chuangqwires}. The particular
electronic band structure of bulk Si leads to fast oscillations in the
electronic exchange coupling when the interacting donor-pair
relative position is changed on a lattice-parameter
scale~\cite{Belita1}. Thus,  proposals based on this mechanism
require nanofabrication techniques far beyond current capabilities.
\begin{figure}[t]
\begin{center}
 \resizebox{!}{7.5cm}{\includegraphics*{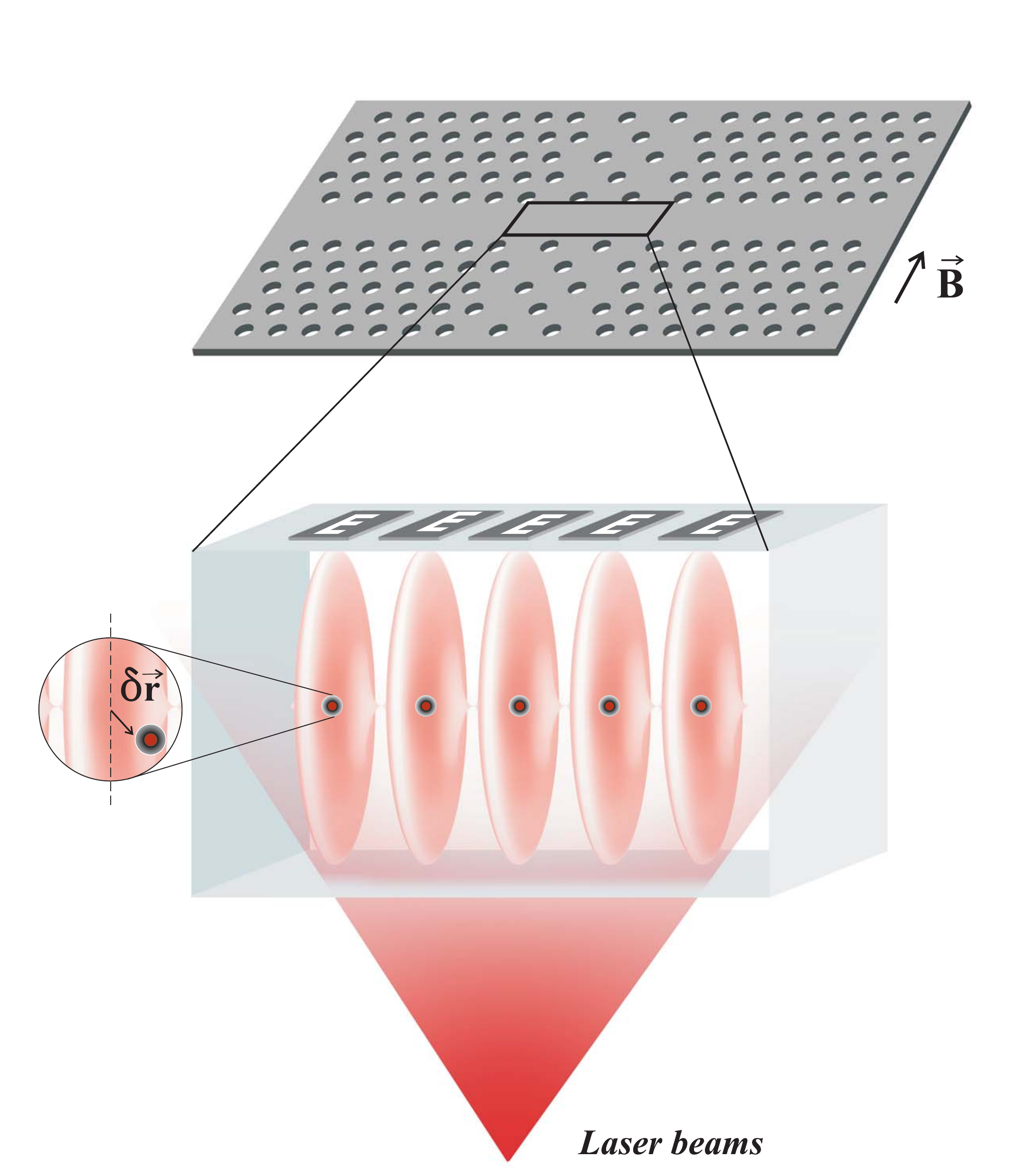}}
\caption{{\bf Proposed quantum-computer architecture.} Donor
impurities are placed in the neighborhood of intensity maxima of a photonic crystal
cavity mode, not necessarily every maximum. The donors are under the action of a uniform magnetic
field $\mbf B$ and electric fields $\mbf E$,   produced by the
electrodes $E$. The
magnetic field is strong enough to decouple nuclear and electronic spins in the donor ground state, and Zeeman-splits two electronic spin states,
$|\uparrow\ket$ and $|\downarrow\ket$, which constitute the qubit.
Turning on the electric fields allows to switch off individual
qubit Raman transitions induced by the two laser beams,
  spread out over the ensemble of qubits, and
  also the coupling among different qubits through the vacuum cavity field.
  The inset displays the misplacement $\delta\vec r$ of an impurity from
  a maximum of the cavity mode.} \label{computer}
\end{center}
\end{figure}

Here we propose placing the donors in a single-mode photonic-crystal
Si cavity \cite{nanocavity}, and optically addressing them through Raman
transitions induced by the cavity mode and only three applied laser
beams, spread out over the whole ensemble. A fourth laser beam is
used for the readout. The qubits result from the interaction of the
electron spin with a uniform magnetic field ($B$). The system is kept at a temperature around 7 K. Essential
elements of the architecture considered here are schematically
illustrated in Fig.~\ref{computer}, where we represent an array of
donors positioned at the maxima of the cavity mode. One- and
two-qubit logical gates, as well as system initialization and
readout, are implemented through the external laser beams, which
address all qubits simultaneously. The coupling between a donor
electron and the light fields may be interrupted by an external
electric field, due to the Stark shift of the donor levels. This
effect is explored for selecting the target qubits for one- or
two-qubit operations. Two-qubit operations are mediated by the
vacuum field of the photonic-crystal cavity~\cite{Imamoglu}.

\begin{figure}[t]
\begin{center}
 \resizebox{!}{5cm}{\includegraphics*{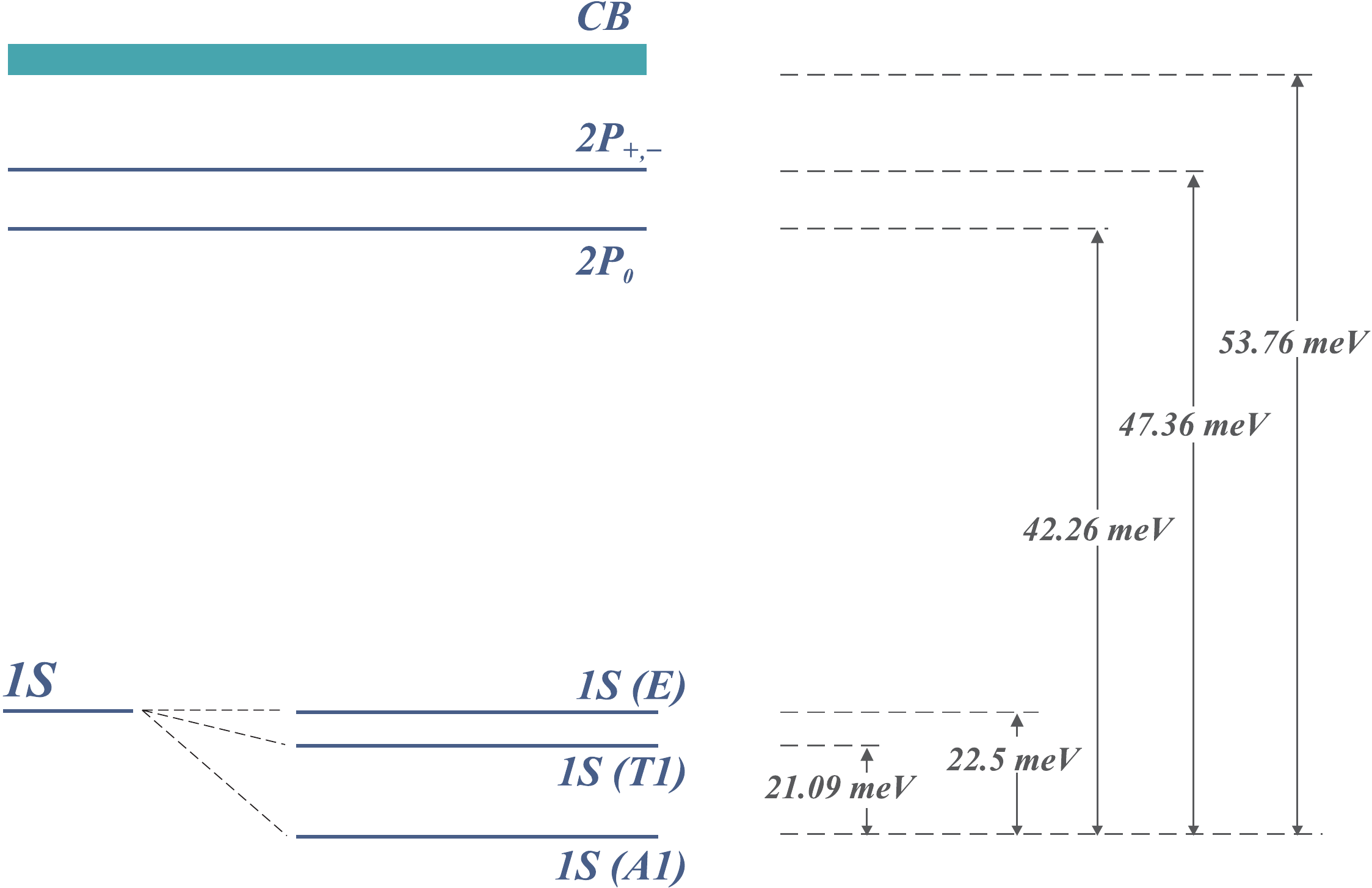}}
\caption{{\bf Orbital energy levels of As in Si.} Only the relevant
levels are shown. Here CB stands for conduction band.}
\label{levels}
\end{center}
\end{figure}

The study of group-V donor impurities in silicon is a quite mature
field~\cite{Kohn,Ramdas} . When a group-V element, such as P, As or
Sb, substitutes the group-IV Si atom in bulk Si, an additional
electron is incorporated in the system. This electron remains bound
to the core potential via a screened Coulomb interaction,
constituting a solid-state analogue of the hydrogen atom.
Electronic-structure peculiarities of bulk silicon lead to a
modified hydogenic spectrum for the donor.
Experimental values for the lowest energy levels and relative energies
for As donors in silicon~\cite{Ramdas} are given in Fig.~\ref{levels}.

Qubit states are defined within the $1S(A1)$ (see Fig.~\ref{levels})
ground state under a magnetic field $B$ satisfying $g_e\mu_{B}B\gg A$, where $A$ is the
hyperfine coupling constant, $\mu_{B}$ is the Bohr magneton and $g_e
(\approx 2)$ is the electron Land\'e factor in Si. For As in Si,
$A=400$ MHz, so a magnetic field $B=0.3$ T will decouple nuclear
and electron spin states, generating well-defined electron spin
states $|\downarrow\rangle$ and $|\uparrow\rangle$ as qubits, with a
10 GHz splitting.

\begin{figure}[b!]
\begin{center}
  \resizebox{8.5cm}{!}{ \includegraphics*{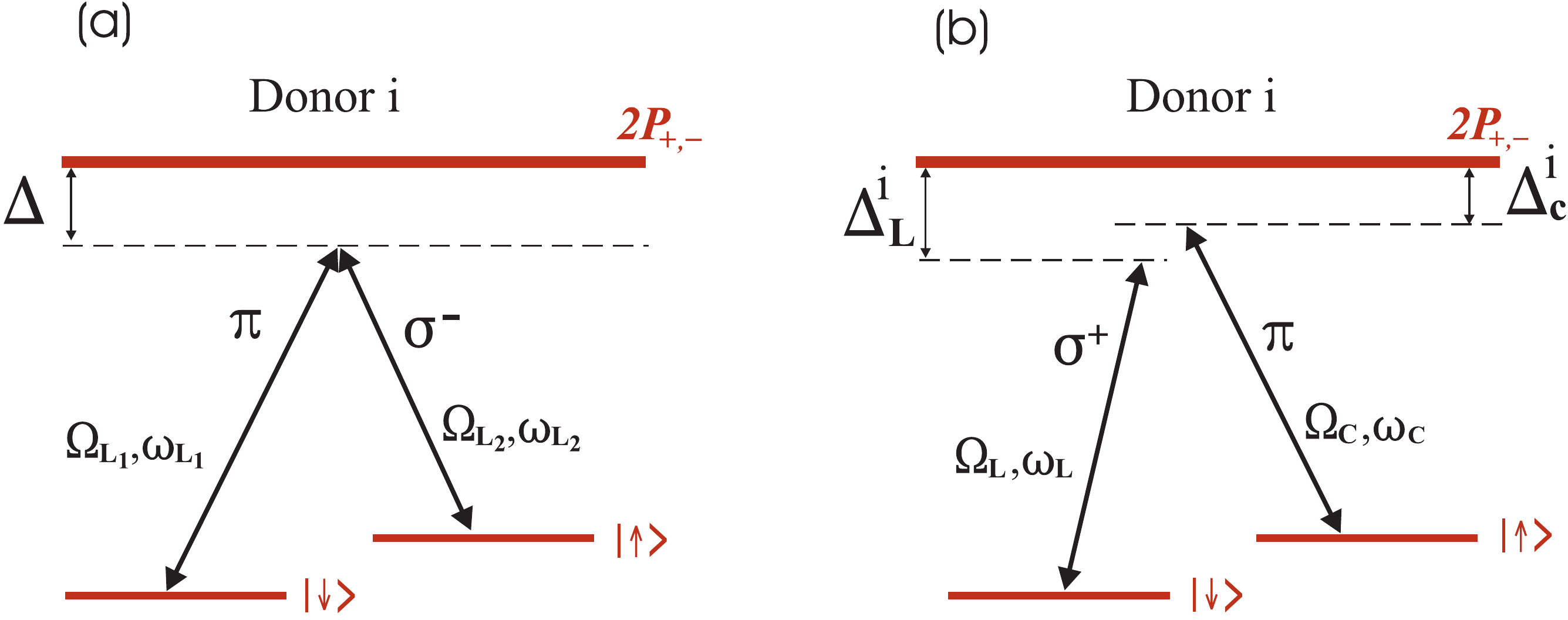}}
\caption{{\bf Logical gates.}
  (a) One-qubit operations are implemented with a linearly polarized
  beam along the direction of the magnetic field and a circularly
  right-polarized beam; (b) Two-qubit operations are implemented
  with a linearly polarized cavity field along the direction of the
  magnetic field and a circularly left-polarized laser beam.}
\label{qbitoperations}
\end{center}
\end{figure}

One-qubit operations are implemented through Raman coupling of the
states $|\downarrow\rangle$ and $|\uparrow\rangle$ of donors
previously selected by switching off the Stark-shift electric fields acting on them. In this excitation
scheme, the donors interact with two laser fields of Rabi
frequencies $\Omega_{L_11}$ and $\Omega_{L_2}$, and frequencies
$\omega_{L_1}$ and $\omega_{L_2}$, respectively, detuned from the
transitions between the qubit states $|\downarrow\rangle$ and
$|\uparrow\rangle$ and the states $|2P_{+,-}\rangle$ by $\Delta$, as
shown in Fig.~\ref{qbitoperations}(a) (the fine structure
splitting of the levels $2P_{+,-}$ is not shown). If
$\Delta\gg\Omega_{L_1},\Omega_{L_2},\Gamma_{p}$, with $\Gamma_{p}$
being the decay rate of the levels $2P_{+,-}$, the levels $2P_{+,-}$
are only virtually populated, giving rise to an effective coupling
between levels $|\downarrow\rangle$ and $|\uparrow\rangle$,
described, in the interaction picture, if
$\Omega_{L_1}=\Omega_{L_2}$, by the Hamiltonian $\widehat{H}_{\rm
eff}=\hbar\Omega_{\rm
  eff}\,|\downarrow\rangle\langle\uparrow| + H.c.,$
where $\Omega_{\rm eff}=\left|\Omega_{L_1}\right|^2/{\Delta}$.

During the Raman coupling of the qubit states there is a small
probability, of the order $\Omega_{\rm eff}/\Delta$, of
populating the intermediate level $2P_{+,-}$. This will lead to  decoherence of the
one-qubit operations with the rate $\Omega_{\rm
eff}(\Gamma_p/\Delta)$ due to the spontaneous decay of the level
$2P_{+,-}$. Since the time required for one-qubit operations is of
the order of $1/\Omega_{\rm eff}$, the error probability per gate
will be $\epsilon_1\approx\Gamma_p/\Delta$.

We propose a spin-orbit (SO) mediated coupling for the
opposite-spin qubit states.
In order to avoid destructive interference effects that would make
$\Omega_{\rm eff}$ vanishingly small, the intensity $\zeta$ of the
SO coupling among the states within the $2P_{+,-}$ manifold
must be comparable or larger than the detuning $\Delta$ of the laser fields.
Strong SO splittings have been
measured for states of the fundamental manifold in the group VI
donors Se:Si and Te:Si, and of their corresponding ionized
states~\cite{socoupling1,socoupling2,socoupling3}.
For Se:Si, the measured SO coupling is 3.2 cm$^{-1}$ (96 GHz);
so we assume in our calculations for As, the element corresponding to the
Se row in the periodic table,  a coupling of the same order of magnitude, $\zeta\sim$ 100 GHz.

The SO coupling also mediates an efficient mechanism for selectively
turning on and off the interaction between the qubits and the light
fields through an applied electric field produced by the electrode above each donor.
The electric field has two effects: It increases the detuning between the Raman laser fields and the atomic transition, so that it becomes much higher than the SO splitting, and it mixes 2P and 2S states. The increase of the detuning causes a destructive interference between the 2P states, which leads to the vanishing of the Raman transition. The mixing of 2P and 2S states also reduces the Raman coupling, since the 2S state does not couple with the ground state. Perturbation theory indicates that the combined effect reduces the transition probability by a factor equal to the third power of the ratio between the SO coupling and the electric dipole energy, which leads to an error of the order of $10^{-4}$ for an applied field equal to 20 kV/cm. This field is below the ionization threshold for P in Si \cite{debernardi}, and for As the threshold shoud be even higher since the binding energies are larger.		
Only the donors that are subjected to smaller electric fields will be
affected by the Raman coupling. This scheme has the advantage that only quiescent atoms are subject to large electric fields, so that the essential properties of the active atoms remain unchanged.

One-qubit operations are implemented with a linearly-polarized beam
along the direction of the magnetic field (same polarization as for the cavity
mode) and a circularly
right-polarized beam, as shown in Fig.~\ref{qbitoperations}(a).
Due to selection rules, the cavity mode does not affect one-qubit operations.

Two-qubit operations involve the interaction of a previously
Stark-shift-selected pair of qubits with an additional laser beam and with
the cavity mode. The frequencies and polarizations of the laser and
cavity fields are chosen in such a way that they nearly satisfy the
conditions for Raman coupling of the qubit states
$|\downarrow\rangle$ and $|\uparrow\rangle$ for each
donor~\cite{Imamoglu} (see Fig.~\ref{qbitoperations}(b)). The
wavelength for the transition $1S(A_1)\rightarrow2P_{+,-}$ is 26.39
$\mu$m in vacuum; in Si (dielectric constant $\epsilon=11.4$), the
corresponding value is $\lambda=7.82$ $\mu$m. The wavelengths of the
cavity mode and the laser beams should be around this value.

Under the conditions established for
a Raman transition and in the dispersive regime, i.e.,
$\Delta_L^i\gg\Omega_L,\Gamma_p$ and
$\Delta_C^i\gg\Omega_C,\Gamma_p,\Gamma_C$, where $\Omega_C$ quantifies the coupling with the cavity
mode, with width $\Gamma_C$,
the states $2P_{+,-}$ are only virtually occupied, giving rise to an
effective coupling between the cavity mode and the qubit states
which, in an
adequate interaction picture, is described by the
Hamiltonian:
\begin{equation}
\widehat{H}_{\rm eff}=\sum_i\left[\hbar\Omega_{\rm
    eff}^i\,\hat{a}\,\hat{\sigma}_-^ie^{-i\delta^i t}+ H.c.\right],
\end{equation}
where $\hat{a}$ is the annihilation operator for the
cavity field,
$\sigma_-^i=|\downarrow\rangle_{i\,i}\!\langle\uparrow|$ is the spin
flip operator for donor $i$, $\delta^i=\Delta_L^i-\Delta_C^i$,
and the sum extends over all the donors selected by the electric
static fields. The couplings $\Omega_{\rm eff}^i$ are defined as:
\begin{equation}
\Omega_{\rm eff}^i=\frac{1}{2}\Omega_L\Omega_C^*\left(\frac{1}{\Delta_L^i}+
  \frac{1}{\Delta_C^i}\right).
\end{equation}

If $\delta_i\gg\Omega_{\rm eff}^i,\Gamma_C$, the cavity field will be only virtually excited and can be
eliminated from the dynamics, leading to an effective two-qubit
interaction mediated by the vacuum of the cavity mode:
 \begin{equation}
\widehat{\widetilde{H}}_{\rm ij}=\sum_{i\ne j}\left[\hbar\Omega_{\rm
    ij} \,\hat{\sigma}_+^i\,\hat{\sigma}_-^je^{i\delta^{\rm ij} t}
+ H.c.\right],
\label{hij}
\end{equation}
where $\delta^{\rm ij}=\delta^i\!-\!\delta^j$. From Eq.~(\ref{hij}),
one can see that each pair of qubits $i,j$ that
satisfies $\delta^i=\delta^j$ will resonantly interact through the
Hamiltonian
 \begin{equation}
\widehat{H}_{\rm ij}=\hbar\Omega_{\rm
    ij} \,\hat{\sigma}_+^i\,\hat{\sigma}_-^j+ H.c.,\label{2qbithamilton}
\end{equation}
with the effective coupling constant $\Omega_{\rm
ij}=\left[\Omega_{\rm eff}^i\left(\Omega_{\rm
      eff}^j\right)^*\right]/{\delta^i}$.
The qubit pairs for which $\delta^{\rm ij}\gg\Omega_{\rm ij}$ will
interact off-resonantly and will not couple to each
other. The error probability per gate for two-qubit operations can
be found in a similar way as for one-qubit operations and will be
given by $\epsilon_2\approx \Gamma_C/\delta^i$.

The Hamiltonian~(\ref{2qbithamilton}) implements the  $\sqrt{\rm
  SWAP}$ operation
$|\uparrow\downarrow\ket\rightarrow(|\uparrow\downarrow\ket
+|\downarrow\uparrow\ket)/\sqrt{2}$, which, combined with
single-qubit rotations, can
be used to implement a CNOT gate~\cite{Loss}. $\sqrt{\rm SWAP}$
operations can be implemented in parallel, by having different
pairs, with $\delta^i=\delta^j$, $\delta^k=\delta^l$, but
$|\delta^i-\delta^k|\gg|\Omega_{ij}|$.

\begin{figure}[t!]
\begin{center}
\resizebox{7.5cm}{!}{ \includegraphics*{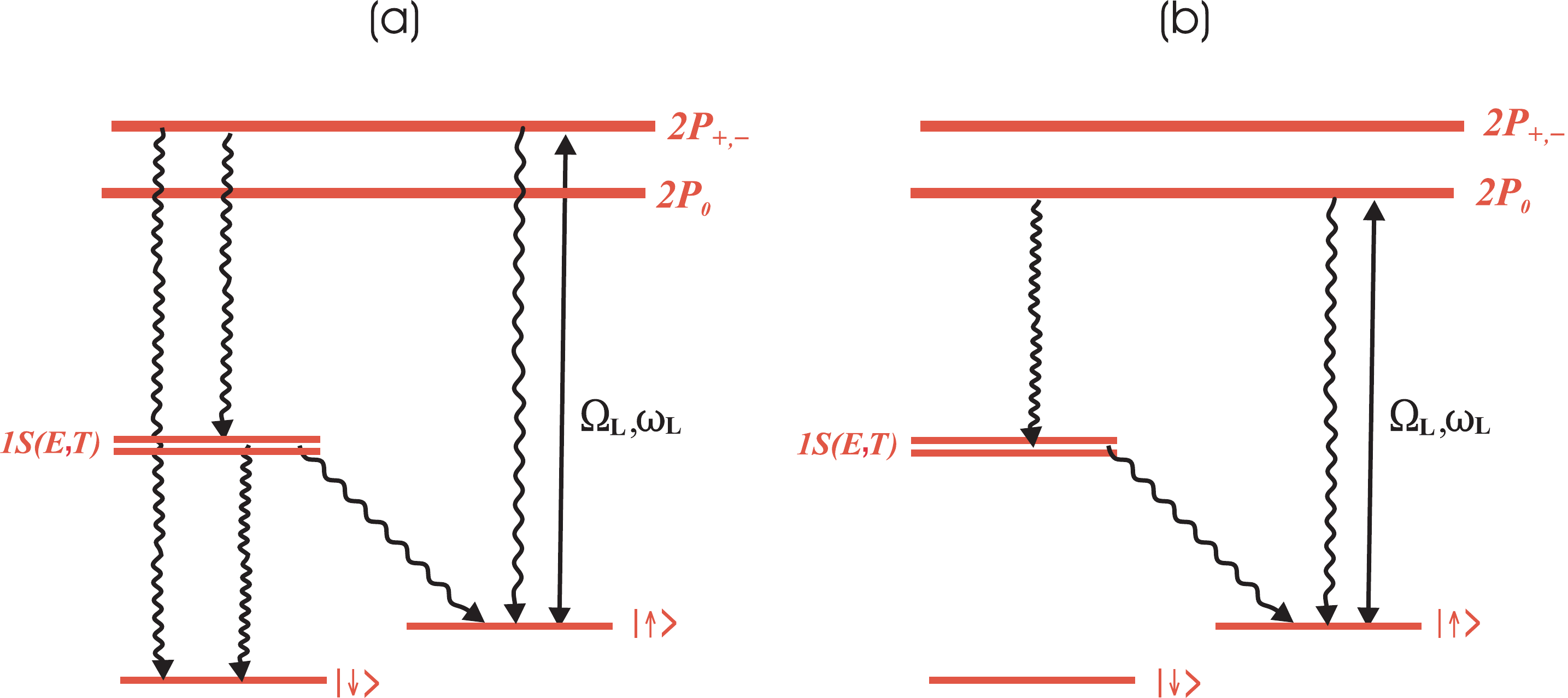}}
\caption{{\bf Initialization and readout.} (a) Initialization:
population of qubit state $|\uparrow\rangle$ is transferred to the
qubit state $|\downarrow\rangle$ via optical pumping by resonantly
exciting transition $|\uparrow\rangle\leftrightarrow 2P_{+,-}$ with
laser light; (b)Readout of the qubit-state is made by monitoring the
fluorescence light of the cyclic transition
$|\uparrow\rangle\leftrightarrow 2P_0$. } \label{initialization}
\end{center}
\end{figure}

The qubits are initialized by driving resonantly the transition
$|\uparrow\rangle\leftrightarrow 2P_{+,-}$ with a laser beam. Under
the action of the magnetic field $B$ and at temperatures of the
order of $7$  K only the state $1S(A1)$ will be populated. Since
the level $2P_{+,-}$ is unstable, it will decay to one of the qubit
levels, leading to optical pumping of the level $|\downarrow\rangle$
(see Fig.~\ref{initialization}(a)).

Qubit readout takes advantage of the fact that the states of the
$2P_0$ manifold do not show SO coupling. If laser light excites
resonantly the transition $|\uparrow\rangle\leftrightarrow 2P_0$,
only the states of that manifold with the same electronic spin as
the state $|\uparrow\rangle$ are excited. Due to selection rules,
the radiative- or phonon-assisted decay of these states to states
with different electronic spin is forbidden. The decay out of the
$2P_0$ level is radiative, whereas the decay out of level $1S(E)$ is
phonon assisted. For this reason, all the excitation will decay back
to the state $|\uparrow\rangle$.  Therefore the transition
$|\uparrow\rangle\leftrightarrow 2P_0$ is cyclic and the electron
shelving technique can be used to measure the occupation of the
qubit states~\cite{leibfried}: If, during the laser excitation of
the transition $|\uparrow\rangle\leftrightarrow 2P_0$ fluorescence
light is observed, the qubit was in state $|\uparrow\rangle$,
otherwise the state $|\downarrow\rangle$ was occupied (see Fig.~\ref{initialization}(b)). Since the
decay $1S(E)\rightarrow|\uparrow\rangle$ is assisted by acoustic
phonons, part of the fluorescence light differs in frequency from
the laser exciting the transition
$|\uparrow\rangle\leftrightarrow2P_0$, which implies that it is
possible to distinguish the fluorescence signal from scattered laser
radiation.

The feasibility of the proposed scheme is based on the following
estimates for the frequencies, couplings, and times involved in the
one-qubit, two-qubit and readout operations.

The measured absorption linewidth of the 1S(A1) $\to 2P_{+,-}$ transition
for Si:P is approximately 1GHz~\cite{Cardona}, giving an upper bound for the decay rate
$\Gamma_p$. This rate could be strongly decreased (more than one order of magnitude), since
 phonon-mediated decay can be suppressed by applying stress, as demonstrated
in Ref.~\cite{pavlov1}, and
spontaneous radiative transitions from these levels are also strongly
supressed due to the photonic band gap, since they are far detuned
from the cavity mode of the photonic crystal~\cite{imamoglu2}.  The Raman-coupling conditions are satisfied, for example, for
$\Omega_{L_1}/2\pi=\Omega_{L_2}/2\pi=2{\rm GHz}$ and $\Delta=200{\rm
GHz}$. This would lead to an effective Rabi frequency $\Omega_{\rm
eff}/2\pi=20{\rm MHz}$ and an error probability per gate
$\epsilon_1\approx 10^{-4}$. For these parameters, which correspond to laser powers of the order of $10$ mW, the
typical time for a one-qubit operation would be of the order of
$50\,{\rm ns}$, much shorter than a spin decoherence time of
$60\,{\rm ms}$. Under the same conditions, we have calculated that the error probability per gate induced by eventual impurity ionization, due to two-photon absorption, is negligibly small ($\epsilon\approx 6\times10^{-7}$).

For the two-qubit operations, one could choose, for example,
$\Omega_C^i/2\pi\sim 30{\rm MHz}$ (this value corresponds to a modal
volume $100\lambda^3$), $\Omega_L^i/2\pi\sim 5{\rm GHz}$,
$\Delta_L^i=100{\rm GHz}$ and $\Delta_C^i=99{\rm GHZ}$. This yields
an effective two-qubit coupling $\Omega_{\rm ij}/2\pi\sim 2.25{\rm
KHz}$, which allows one to perform more than $10^3$ $\sqrt{\rm
SWAP}$ operations within a qubit decoherence time of 60 ms. Since
for two-qubit operations the error per gate is
$\epsilon_2\approx\Gamma_C/\delta_ i$, an error per gate of the
order of $1\times 10^{-3}$ would imply a cavity decay rate
$\Gamma_C\approx 1{\rm MHz}$. This requires a cavity quality factor
$Q\approx 10^7$. Quality factors of $10^6$ have already been
reported for silicon-based photonic-crystal nanocavities
\cite{nanocavity}; $Q$'s as high as $2\times 10^7$ seem to be within
reach \cite{nanocavity}. Larger wavelengths in the $\mu$m region, as
used in our proposal, should lead to yet higher values of $Q$.
Combined with larger values of the spin decoherence time, consistent
with the experimental results \cite{tspin,dassarma}, this would
allow one to increase $\delta$, further reducing the error per gate.

Readout is very fast, since the decay rate out of level $P_0$ is of
the order of 1 GHz whereas the phonon-assisted decay of level
$1S(E)$ is of the order of $10^{-10}$ s~\cite{pavlov}. We estimate
that about 10000 cycles, corresponding to a detection time around 1
$\mu$s, should yield a reading efficiency close to 100\%.
Parallel readout can be implemented for donors separated by ten
cavity wavelengths or more.

Finally, we address a crucial fabrication issue: Given that the best
currently achievable deposition control for impurities in Si is
$\sim10$ \AA~\cite{Oberbeck}, the impact of small donor
misplacements on the proposed device operation must be analyzed. A
deviation $\delta\vec r$ in the position of a donor from a maximum
of the cavity field (see Fig.~\ref{computer}) introduces a variation
$\Delta \Omega_C\approx 2\pi^2(|\delta\vec r|/\lambda)^2\Omega_C$ on the
cavity vacuum Rabi frequency $\Omega_C$ at the position of the
donor. Here $\lambda=7.8\,{\rm\mu}$m is the cavity wavelength. This
implies that
$|\delta\vec r|=100 \textrm{\AA}$ leads to 
$\Delta\Omega_C\approx 3\times 10^{-5}\Omega_C$. The time for a
typical two-qubit gate such as $\sqrt{\rm SWAP}$ is $t\sim
1/\Omega_{\rm ij}$, leading to an error probability in this
operation of $p\approx (\Delta\Omega_C/\Omega_C)^2\approx 1\times
10^{-9}$, which means that our operation scheme is quite insensitive
to relatively large (several lattice parameters) donor misplacement
within the simple donor linear array architecture. This is in
contrast with Kane's original exchange-based proposal, which leads
to much more stringent conditions on impurity positioning, and
requires elaborate two-dimensional architectures to compensate for
donor misplacement~\cite{Hollemberg}.

Our estimations indicate that the present proposal could meet the
conditions for a robust quantum computation device. The precise
quantum control of atoms and ions in optical cavities, already
demonstrated in several labs, and the fact that Si is the leading
material in terms of processing and device fabrication, with
sophisticated techniques for impurity implantation and high-$Q$
microcavity construction, indicate that this system might be a viable candidate
for a working quantum computer using achievable
technological resources.

\begin{acknowledgments}
  The authors acknowledge financial support from the Brazilian funding
agencies CNPq, CAPES, PRONEX, FUJB and FAPERJ. This work was
performed as part of the Brazilian Millennium Institutes for Quantum
Information and Nanotechnology.
\end{acknowledgments}

\end{document}